\documentclass[conference]{IEEEtran}
\IEEEoverridecommandlockouts
\usepackage{comment}
\usepackage{cite}
\usepackage{amsmath,amssymb,amsfonts}
\usepackage{algorithmic}
\usepackage{graphicx}
\usepackage{textcomp}
\usepackage{float}
\usepackage{xcolor}
\usepackage{amsmath}
\usepackage{subfig}
\usepackage[normalem]{ulem}
\usepackage{soul}
\usepackage[mathletters]{ucs}
\usepackage[utf8x]{inputenc}

\def\BibTeX{{\rm B\kern-.05em{\sc i\kern-.025em b}\kern-.08em
    T\kern-.1667em\lower.7ex\hbox{E}\kern-.125emX}}
\begin{document}

\title{Mobility Based SIR Model For Pandemics -- With Case Study Of COVID-19}

\author{\IEEEauthorblockN{Rahul Goel}
\IEEEauthorblockA{
\textit{Institute of Computer Science}\\
University of Tartu, Estonia \\
rahul.goel@ut.ee}
\and
\IEEEauthorblockN{Rajesh Sharma}
\IEEEauthorblockA{\textit{Institute of Computer Science} \\
University of Tartu,Estonia\\
rajesh.sharma@ut.ee}
}

\maketitle

\begin{abstract}  
In the last decade, humanity has faced many different pandemics such as SARS, H1N1, and presently novel coronavirus (COVID-19). On one side, scientists are focusing on vaccinations, and on the other side, there is a need to propose models that can help us in understanding the spread of these pandemics as it can help governmental and other concerned agencies to be well prepared, especially from pandemics, which spreads faster like COVID-19. The main reason for some epidemic turning into pandemics is the connectivity among different regions of the world, which makes it easier to affect a wider geographical area, often worldwide. In addition, the population distribution and social coherence in the different regions of the world is non-uniform. Thus, once the epidemic enters a region, then the local population distribution plays an important role. Inspired by these ideas, we proposed a mobility-based SIR model for epidemics, which especially takes into account pandemic situations. To the best of our knowledge, this model is first of its kind, which takes into account the population distribution and connectivity of different geographic locations across the globe. In addition to presenting the mathematical proof of our model, we have performed extensive simulations using synthetic data to demonstrate our model's generalizability. To demonstrate the wider scope of our model, we used our model to forecast the COVID-19 cases for Estonia.

\end{abstract}


\section{Introduction}\label{sec:Intro}
In this modern age, pandemics are not a rare phenomenon. As in the last decade, we have seen several pandemics such as H1N1, SARS, EBOLA, and presently in 2020 humanity is facing its biggest crisis due to COVID-19. The severity of these pandemics can be understood by the death toll claimed by them. According to WHO, the pandemic H1N1/09 virus resulted in 18,036 deaths \cite{world2009pandemic}. On the other hand, the CDC estimate between 151,700 to 575,400 deaths due to the pandemic H1N1/09 virus ~\cite{centersfirst}. Currently, the coronavirus (COVID-19) pandemic, which started in December 2019 from Wuhan, China has infected 2,404,249 individuals and claimed 165,229 (as of $20^{th}$ April 2020) deaths worldwide ~\cite{csse2020coronavirus} ~\cite{covid19global}. Pandemics are different from epidemics in terms of their geographic spread. An \textit{epidemic} affects many people at the same time. It spreads from person to person and remains local to a specific region. In comparison, when an epidemic  engulfs an entire country, continent, or the whole world, it is termed as \textit{pandemic}.

In the past, various models have been proposed for understanding the epidemic spreads. These models can be broadly classified into two categories, that is agent-based modeling \cite{bonabeau2002agent}~\cite{schelling1971dynamic}~\cite{sun2006cognition} and compartmental models \cite{kermack1927contribution}~\cite{hethcote2000mathematics}~\cite{goel2019modeling}. The agent-based modeling is used for simulating the actions and interactions of autonomous agents as a whole \cite{epstein2009modelling}. These agents can be both individual or collective entities such as organizations or groups. In contrast, differential equations are used in compartmental models, where the population is divided into different compartments such as suspected (S), infected (I), and recovered (R) ~\cite{kermack1927contribution}. Several other variants of these models have also been proposed such as SI ~\cite{hurley2006basic}, SIS ~\cite{naasell1996quasi}, SIR ~\cite{kermack1927contribution}, SIRS 
~\cite{jin2007sirs}, etc.

\begin{figure}
 \centering
 {\includegraphics[width = 0.9\columnwidth]{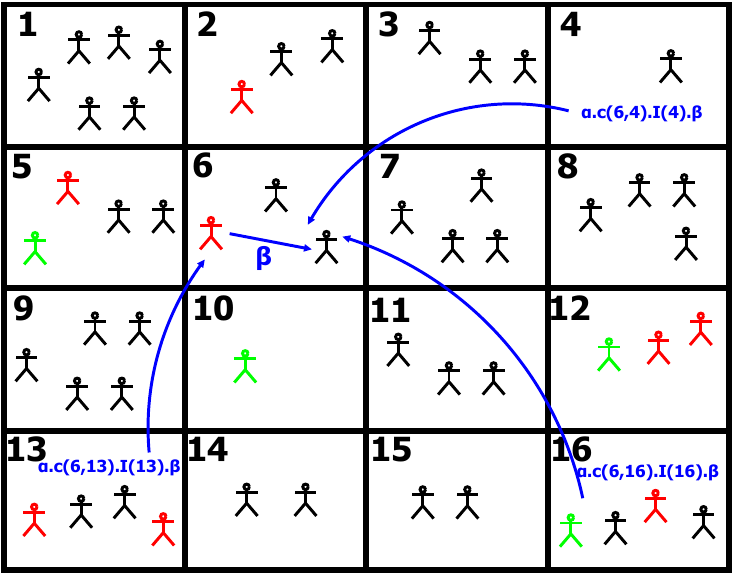}}
 \caption{\textbf{Local And Global Transmission Of Infection 
 }: Each cell represents 
 a separate region with some population density (here regions are 1 to 16). Individuals in each cell are color-coded: Black (Susceptible), Red (Infected) and Green (Recovered). The local transmission rate of infection is $\beta$ for all cells. For region 6, it's social connectivity with other regions is $\alpha$. The mobility of individuals from region 4 to region 6 and fraction of infected individual at region 4 are represented as $c(6,4)$ and $I(4)$ respectively. Therefore, infection can transfer from region 4 to 6 via global transmission rate $\alpha c(6,4) I(4) \beta$. Similarly, $\alpha c(6,13) I(13) \beta$ and $\alpha c(6,16) I(16) \beta$ signifies the global transmission rate from region 13 and 16 respectively to region 6.}
 \label{fig:method}
\end{figure}

Compartmental models are often being criticized by the agent-based model researchers because they struggle to capture the connectivity between different regions of the globe, and different real-world population characteristic, such as worldwide population distribution 
~\cite{chinazzi2020effect}~\cite{eubank2004modelling}. In this study, we proposed a mobility-based model, an extension to the classical SIR based epidemic model, which considers the real-world population distribution across different regions of the world. Most importantly, the model also takes into account the connectivity factor among various regions of the world, which is the key cause in accelerating the process of transforming epidemics into pandemics. We model the regions in a 2-dimensional lattice, where each cell represents the mobility parameter (or direct connectivity) from one region to another. Along with presenting the  mathematical proof of our model, 
we have performed extensive simulations on synthetic data and forecast the COVID-19 cases in Estonia\footnote{https://koroonakaart.ee/en} by inferring the interaction among individuals through call data records between Estonian counties to demonstrate the model's ability to generalize on different types of data.

The proposed model is composed of (local) transmission rate of the infection $\beta$ and to cover the mobility aspect, we introduce parameters: 1) `$\alpha$' which is a social connectivity parameter that signifies how well individuals are socially linked with each others, and 2) `$c_{(i,j)}$' that represents individuals mobility from some region $j$ to another region $i$. Thus, the infection can transfer within the region with the transmission rate $\beta$ and can also be introduced from other regions through global transmission rate which depends upon $\alpha$, $c_{(i,j)}$, $I_j$ (fraction of infected at region j) and $\beta$. With the help of Figure \ref{fig:method}, we illustrate our proposed model for better understanding. We applied our model on synthetic network as well as on a real network of Estonia considering the population density and the connectivity among counties, which is created using call data records (CDR) to investigate the following questions:

\begin{itemize}
    \item \textit{How \textit{social} connectivity parameter `$\alpha$' affects the fraction of individuals in different compartments (\textit{susceptible, infected and recovered}) during a pandemic?} We address this question by carefully examining the effect of $\alpha$ while keeping all the other parameters constant (Section \ref{subsec:results}).
    \item \textit{What are the outcomes of restricting mobility from the top-X percentile of strongly connected regions?} We explore the outcomes of mobility restriction with the model and found that restricting the mobility of top-10 percentile of connected regions can reduce the number of infected individuals between 18\% to 27\% (Section \ref{subsec:results}).
    \item \textit{What is the relationship between \textit{social connectivity parameter} `$\alpha$' and mobility restriction (of top-X percentile) of strongly connected regions?} To address this question, we performed numerical simulation on the proposed mean-field equations (Section \ref{subsec:results}, Figure \ref{Fig:RelationRandom}).
    \item \textit{How efficiently this model can perform in real scenarios?} We answer this question by projecting the expected COVID-19 cases in Estonia using the model and compared the results with the real cases (Section \ref{subsubsec:Estonia}). 
\end{itemize}

The limitation of classical compartmental epidemiological models is that they do not take into account the importance of reducing social connectivity (or isolation) and the significance of mobility restriction during the spreading of a pandemic such as COVID-19. This limitation is overcome in the proposed model. We found that the reproduction number $R_0$ for a pandemic depends upon the social connectivity and mobility parameter. We also discovered that during a pandemic, restricting mobility reduces the fraction of individuals in an infected compartment and reducing the social connectivity (or isolation) delays the peak and also reduces the number of infected individuals from the pandemic. We believe that this model can help to adopt a balanced strategy to address a pandemic crisis.

The rest of the paper is organized as follows. Next, we discuss related works with respect to epidemic modeling. We then describe the model preliminaries and derivations in Section \ref{sec:methodology}. Section \ref{sec:evaluation} presents the evaluation results of our model and we conclude with a discussion of future directions in Section \ref{Sec:Conclusion}.
\section{Related Work}
In this section, we discuss relevant literature with respect to epidemic modeling which involves two different lines of work. First involving agent-based modeling and the second using compartmental based modeling. In the agent-based modeling, authors model epidemics by simulating the actions and interactions of autonomous agents (both individual or collective entities such as organizations or groups) with a view of assessing their effects on the system as a whole~\cite{epstein2009modelling} by using transportation systems such as road networks~\cite{eubank2004modelling}, airways~\cite{chinazzi2020effect} etc. These models have been used for understanding various epidemics such as smallpox~\cite{burke2006individual}, influenza~\cite{khalil2012agent}, cholera~\cite{crooks2014agent}, and very recently about COVID-19~\cite{chinazzi2020effect}.

In contrast to agent-based modeling, a differential equation based compartmental models have also been used for understanding epidemics, which is the basis of this work. This line of literature is mainly based on the classical SIR model proposed by Kermack and McKendrick ~\cite{kermack1927contribution}
followed by ~\cite{anderson1979population} ~\cite{anderson1992infectious}. In \cite{anderson1979population}, the authors considered the host population as a dynamic variable rather than constant, as conventionally assumed, which provides a broader understanding of the population behavior during infectious disease. In their work in ~\cite{anderson1992infectious}, authors discuss the idea of the basic reproductive rate, threshold about host densities, and modes of transmission.

Different variations of SIR model have also been proposed to capture various real-world scenarios. For example, introducing a delay in the model to capture the incubation period during the spreading \cite{zhang2010analysis,xia2012sir, liu2015delayed, arquam2018modelling} or the introduction of interventions such as antiviral drugs ~\cite{towers2011antiviral}. In a different work to represent non-linear nature of epidemic spread, a SIR rumor spreading model was proposed in which tie strengths were dependent on nodes' degree ~\cite{singh2012nonlinear}. Apart from SIR based models, there exist different flavors of compartmental models, which represent different scenarios such as SIS ~\cite{naasell1996quasi}, where individuals do not recover and can become susceptible again. This model has also been studied using varying types of underlying topologies ~\cite{shi2008sis}.

A set of works has also focused on exhibiting the epidemic spreading by using varying types of underlying network structures. For example, authors in ~\cite{moreno2002epidemic}, ~\cite{barthelemy2005dynamical} and ~\cite{vespignani2012modelling} used a scale-free network and in~\cite{li2006controlling} a small-world evolving networks for evaluating their epidemiological framework. In their work in ~\cite{kiskowski2014three}, authors combine a discrete, stochastic SEIR (E stands for exposed) model with a three-scale community network model to demonstrate that the different regional trends may be explained by different community mixing rates. A detailed study with respect to various epidemic models on varying topologies has been done in ~\cite{pastor2015epidemic}.

In another line of work, authors proposed models to understand epidemics based on the speed of growth. For example, in ~\cite{viboud2016generalized}, authors applied their generalized-growth model to characterize the ascending phase of an outbreak on 20 different epidemics. Their findings revealed that sub-exponential growth is a common phenomenon, especially for pathogens that are not airborne. In another work \cite{huang2016bayesian}, researchers explain the rapid spread of H1N1 in 2009 around the world by using a flexible Bayesian, space–time, Susceptible-Infected-Recovered (SIR) modeling approach. \cite{gojovic2009modelling} developed a simulation model of a pandemic (H1N1) 2009 outbreak in a structured population using demographic data from a medium-sized city in Ontario and epidemiologic influenza pandemic data. In comparison to previous works, the proposed model introduces mobility and social connectivity parameters, the key characteristics for turning epidemics into pandemics.

\begin{table}[h!]
\caption{Parameters description}
    \begin{tabular}{|l|l|}
    \hline
    \textbf{Notations} & \textbf{Meaning} \\\hline
    $l$ & Number of locations \\\hline
    $c$ & Connection between locations \\\hline
    $S_i(t)$ & Number of susceptible individual at location $i$ at time $t$\\\hline
    $I_i(t)$ & Number of infected individual at location $i$ at time $t$\\\hline
    $R_i(t)$ & Number of recovered individual at location $i$ at time $t$\\\hline
    $N_i(t)$ & Population at location at time $t$ $i$\\\hline
    $\alpha$ & Social connectivity parameter\\\hline
    $\beta$ & Infection rate\\\hline
    $\mu$ & Recovery rate\\\hline
    $c_{i,j}$ & Individuals mobility from location $j$ to $i$\\\hline
    \end{tabular}
    
    \label{table:Notation}
\end{table}

\section{Model Preliminaries And Derivations} \label{sec:methodology}

In this section, we first explain the classical \textit{SIR} model and then discuss its limitations with respect to the absence of mobility and social connectivity parameters. Next, we describe our proposed model to understand the spreading of an infection during a pandemic. 

In 1926, Kermack and McKendrick~\cite{kermack1927contribution} proposed the classical SIR model as follows:
\begin{eqnarray}
\frac{ds(t)}{dt} & = & -\beta s(t)i(t)   \label{RE1} \\
\frac{di(t)}{dt} & = & \beta s(t)i(t) - \mu i(t)  \label{RE2} \\
\frac{dr(t)}{dt} & = & \mu i(t)   \label{RE3}
\end{eqnarray}
where, $s(t)$, $i(t)$, $r(t)$ is the fraction of susceptible, infected and recovered population at time $t$.
However, the classical SIR epidemic model does not consider the heterogeneity and topology of the real-world network. 
To overcome this limitation, we introduce the \textit{mobility} and \textit{social connectivity parameters} in our proposed model.

Let, `$l$' represents the total number of
locations and `$c$' denotes the connection (or individuals' 
mobility) between locations. 
The propagation of infection at each location is explained as: Each healthy individual can get the infection either from 
an infected individual located in the same location (local transmission) or from an individual visiting from other connected locations (global transmission). 
The local transmission rate of infection is represented by $\beta$ and the recovery rate as $\mu$ 
and, $\beta$ and $\mu$ 
$\in$ [0,1]. 
In the next section, we discuss the local transmission of infection and then the global transmission is discussed in detail in the Section \ref{subsec:GlobalTrans}.
\begin{figure}[ht!]
  \centering
  \subfloat[$\alpha$ = 1]{\label{figur:1}\includegraphics[width=0.5\columnwidth]{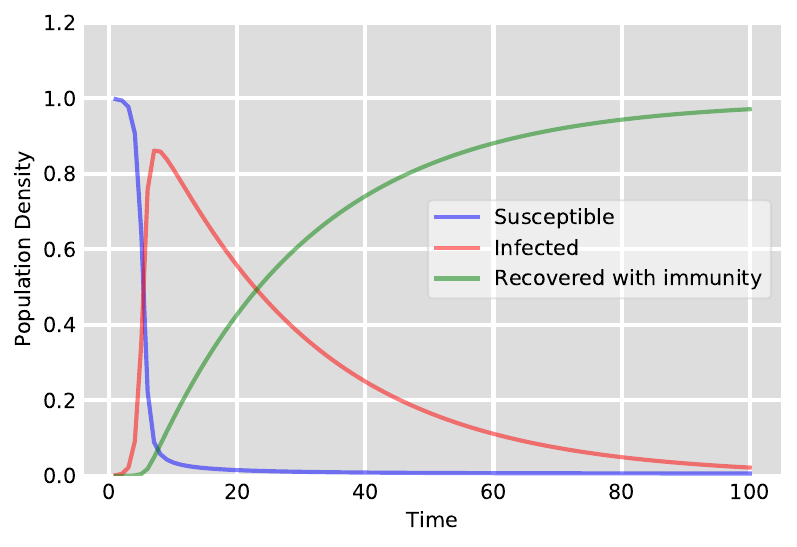}}
  \subfloat[$\alpha$ = 0.8]{\label{figur:2}\includegraphics[width=0.5\columnwidth]{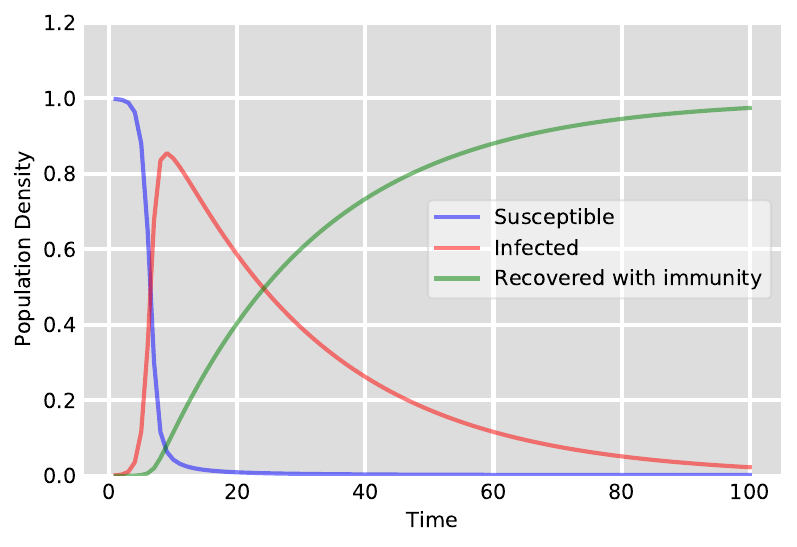}}
  \\
  \subfloat[$\alpha$ =  0.6]{\label{figur:3}\includegraphics[width=0.5\columnwidth]{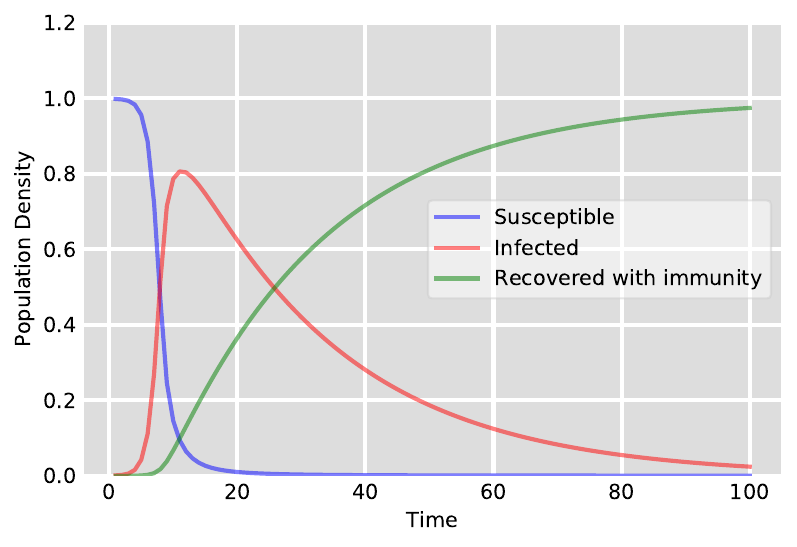}}
  \subfloat[$\alpha$ =  0.4]{\label{figur:4}\includegraphics[width=0.5\columnwidth]{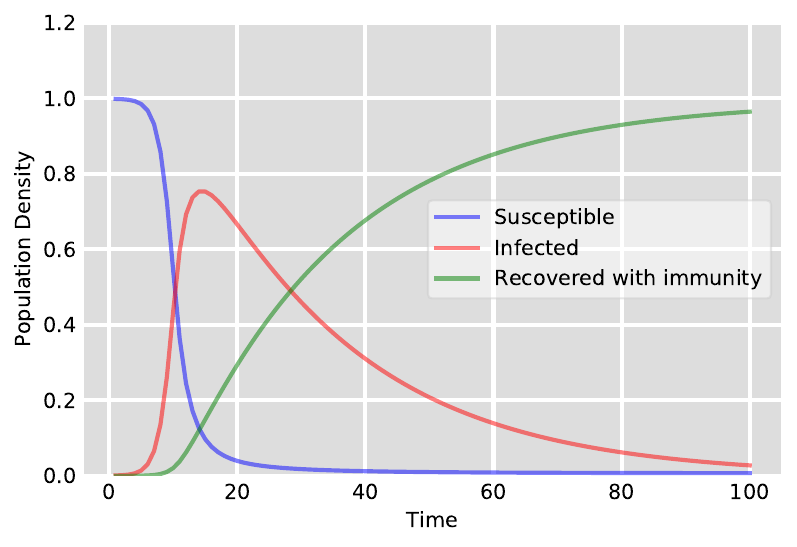}}
  \\
  \subfloat[$\alpha$ =  0.2]{\label{figur:5}\includegraphics[width=0.5\columnwidth]{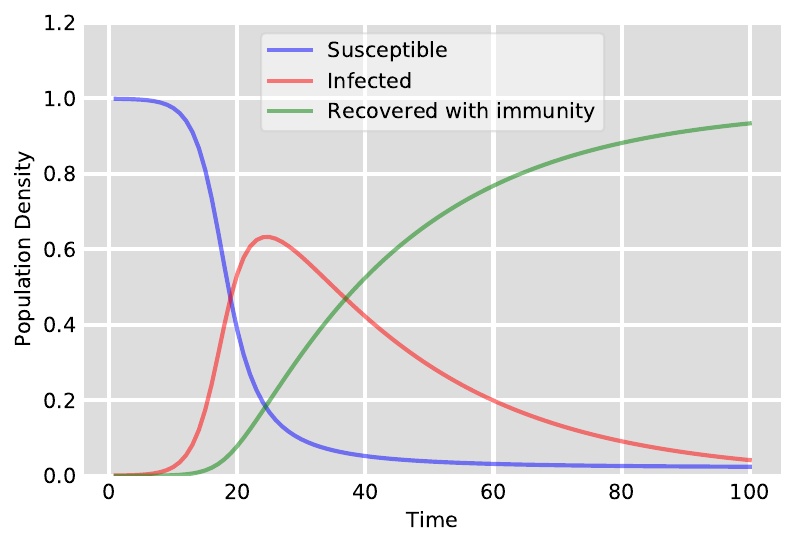}}
  \subfloat[$\alpha$ =  0.1]{\label{figur:6}\includegraphics[width=0.5\columnwidth]{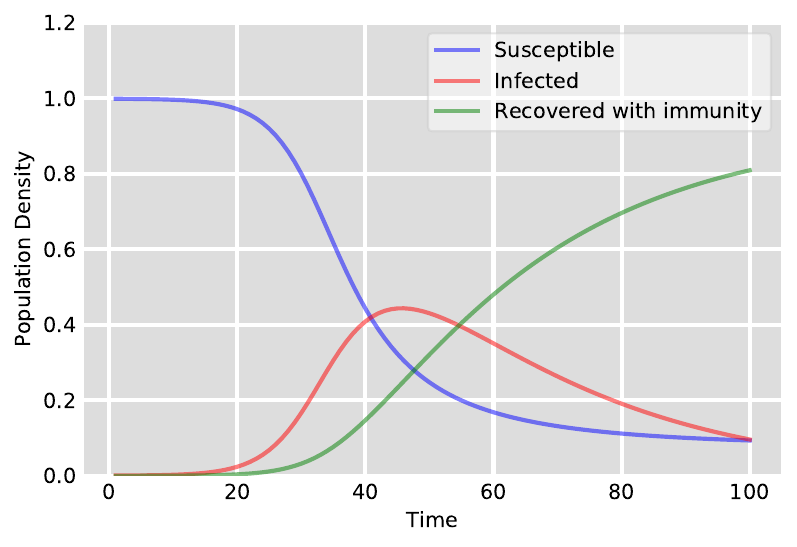}}
\caption{Pandemic Origin From Random Location: Effect of \textit{Social Connectivity Parameter} `$\alpha$'}
\label{Fig:AlphaRandom}
\end{figure}

\begin{figure}[ht!]
  \centering
  \subfloat[0\% Locations Quarantine]{\label{figur:7}\includegraphics[width=0.5\columnwidth]{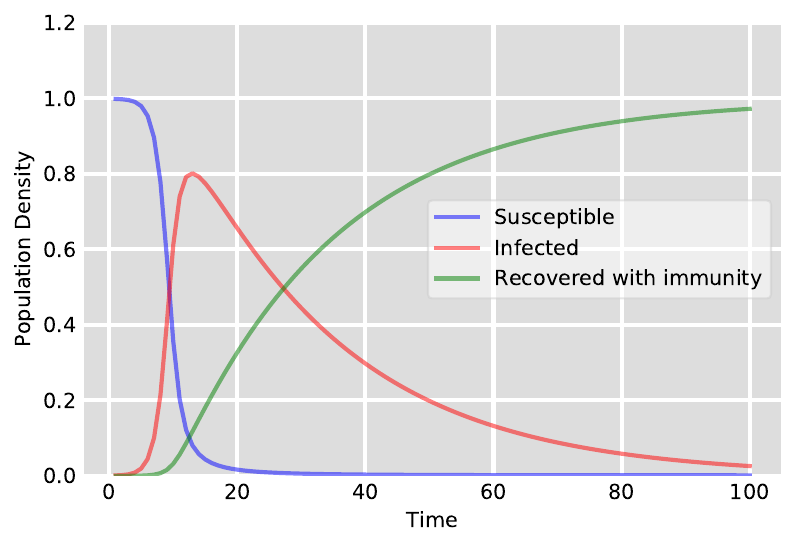}}
  \subfloat[10\% Locations Quarantine]{\label{figur:8}\includegraphics[width=0.5\columnwidth]{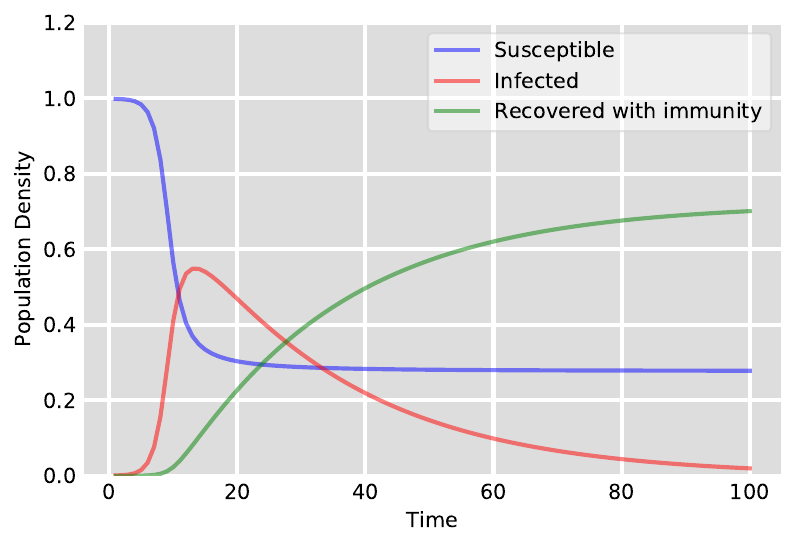}}
  \\
  \subfloat[20\% Locations Quarantine]{\label{figur:9}\includegraphics[width=0.5\columnwidth]{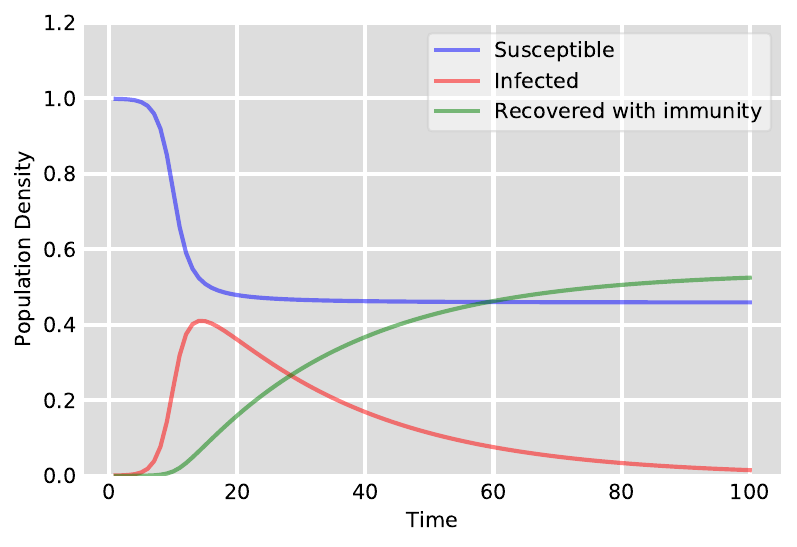}}
  \subfloat[30\% Locations Quarantine]{\label{figur:10}\includegraphics[width=0.5\columnwidth]{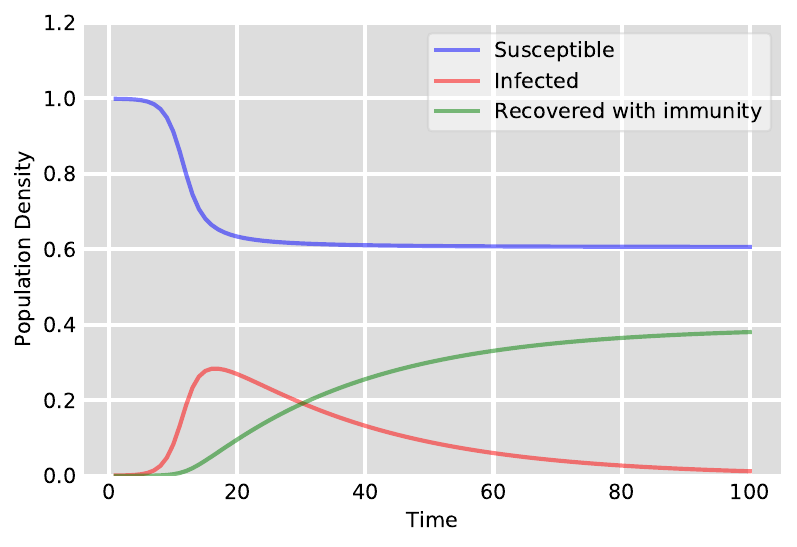}}
\caption{Pandemic Origin From Random Location: Effect of Quarantine Strongly Connected Locations}
\label{Fig:QuantineRandom}
\end{figure}

\subsection{Local Transmission}
Let, $N_i$ 
be the population at location \textit{i}, where $i\in l$ and 
the total population is divided into three compartments. 
The compartments for location \textit{i} at time \textit{t} are as follows:

\begin{enumerate}
    \item $S_i(t)$: the number of individuals susceptible or not yet infected. This compartment is 
    referred as \textit{susceptible compartment}.
    \item $I_i(t)$: the number of 
    infected individuals which can further spread the disease to the 
    individuals present in the 
    susceptible compartment. This compartment is 
    referred to as \textit{infected compartment}. 
     \item $R_i(t)$: the number of individuals who have been recovered from infected compartment. This compartment is 
     referred as \textit{recovered compartment}.
\end{enumerate}

Our assumptions regarding the transmission 
of an individual 
from one compartment to another compartment are 
as follows: 
\begin{enumerate}
    \item A healthy individual after becoming infected moves from susceptible to the infected compartment.
    \item An individual can recovered spontaneously at any time with the recovery rate $\mu$. The recovery of an individual is independent of healthy and infected compartment individuals.
    \item Once the individual 
    gets recovered, it will 
    become immune to the disease and thus, will not transmit the infection to individuals in the susceptible compartment. 
    \item In addition, this model ignores the demography that is birth or death of individuals. Therefore, the population remains constant.
\end{enumerate}

\subsection{Global Transmission}\label{subsec:GlobalTrans}
Let, $j$ ($j$ $\subset$ $l$) represents a set of locations, which 
are connected to location $i$. 
Therefore, $\sum_j N_j$ 
is the maximum possible number of individuals 
connected to location $i$, from all the locations $j$. 
The parameter $c_{i,j}$ reflects the mobility of individuals from locations $j$ to location $i$. Global transmission depends upon 
this mobility parameter 
of individuals from one location to another. 
Similar to local transmission, $I_j$ is the number of individuals in infected compartment 
in all the locations $j$. Hence, total mobility of infected individuals 
from 
all the other connected locations to location $i$ is $\sum_j c_{i,j} \frac{I_j}{N_j}$. 

Considering the above description, the 
chances of transmission of infection from 
all the connected locations to location $i$ is 
$\sum_j c_{i,j} \frac{I_j}{N_j} \beta$. This transmission further depends upon the \textit{social connectivity} ($\alpha$) of individuals at location $i$. Therefore, the proportion of healthy individuals at location $i$ 
which can get infected from infected individuals from location $j$ is $\frac{\alpha \sum_j c_{i,j} \frac{I_j}{N_j} \beta}{N_i + \sum_j c_{i,j}}$. 
Thus, 
the mean-field equations 
for the dynamics of the pandemic, based on the above discussed interactions: 

\begin{eqnarray}
\frac{dS_i(t)}{dt} &=& -\frac{\beta S_i(t) I_i(t)}{N_i(t)} - \frac{\alpha S_i(t) \sum_j c_{i,j} \frac{I_j(t)}{N_j(t)} \beta}{N_i(t) + \sum_j c_{i,j}}\label{eq:S}
\\
\frac{dI_i(t)}{dt} &=& \frac{\beta S_i(t) I_i(t)}{N_i(t)} + \frac{\alpha S_i(t) \sum_j c_{i,j} \frac{I_j(t)}{N_j(t)} \beta}{N_i(t) + \sum_j c_{i,j}} \nonumber\\
&& -\> \frac{\mu I_i(t)}{N_i(t)}\label{eq:I}
\\
\frac{dR_i(t)}{dt} &=& \frac{\mu I_i(t)}{N_i(t)}
\label{eq:R}
\end{eqnarray}

Where, Eq. \ref{eq:S} describes the rate of change of susceptible individuals at location $i$, and Eq. \ref{eq:I} refers to rate of change of infected individuals, and Eq. \ref{eq:R} explains the rate of change of recovered individuals at location $i$. Please refer Table \ref{table:Notation} for notations and their meaning.

\subsection{Dynamical Behaviour Of The Model}
Eq. (\ref{eq:S}-\ref{eq:R}) represents nonlinear dynamical system of pandemic spreading, where at any time $t$,

\begin{eqnarray}
    S_i(t) + I_i(t) + R_i(t) = N_i(t)
\end{eqnarray}

In order to solve mean-field Eq. (\ref{eq:S}-\ref{eq:R}), following assumptions are made (Please note that these assumptions are not considered during our experiments): 
\begin{enumerate}
    \item 
    Initially, the population at all locations is 
    equal to \textit{N(t)} at time $t$.
    \item Individuals in infected compartments are equal to $I(t)$ at all locations at time $t$ and 
    $\sum_j I_j = |j|.I_j = kI_j$, where, $k$ is the number of locations connected to location $i$, that is, $k$ = $|j|$.
    \item The mobility of individuals from one location to another location is a fraction of total population $N$. Let, the sum of fraction of population mobility from $|k|$ locations is $n$. Then, the total individuals mobility from set of locations $j$ to $i$ is $n*N$. Therefore, $\sum_j c_{i,j} = nN$.
\end{enumerate}

By considering the above assumptions, Eq. \ref{eq:S} and \ref{eq:R} can be written as

\begin{eqnarray}
    \frac{dS_i(t)}{dt} &=& -\frac{\beta S_i(t) I(t)}{N(t)} - \frac{\alpha S_i(t) nN(t)k \frac{I(t)}{N(t)} \beta}{N(t) + nN(t)}\label{eq:S_simp}
    \\
\frac{dR_i(t)}{dt} &=& \frac{\mu I(t)}{N(t)}\label{eq:R_simp}
\end{eqnarray}

From Eq. \ref{eq:S_simp} and \ref{eq:R_simp}

\begin{eqnarray}
    \frac{dS_i(t)}{dR_i(t)} &=& -\frac{\beta S_i(t)}{\mu} - \frac{\alpha S_i(t) nk \beta}{\mu(1 + n)}
    \\
    &=& -\frac{\beta S_i(t)}{\mu}\left[1 + \frac{\alpha nk}{1+n}\right]\\
    &=& -\frac{\beta S_i(t)}{\mu}\left[\frac{1+(1+\alpha k) n}{1+n}\right]\label{eq:Solve1}
\end{eqnarray}

For simplicity, Eq. \ref{eq:Solve1} can be written as:
\begin{eqnarray}
    \frac{dS(t)}{dR(t)} &=& -\frac{\beta S(t)}{\mu}\left[\frac{1+(1+\alpha k) n}{1+n}\right]\label{eq:Solve2}
\end{eqnarray}

Eq. \ref{eq:Solve2} can be rewritten as
\begin{eqnarray}
    S &=& S_0 e^{-\frac{\beta}{\mu}R\left[\frac{1+(1+\alpha k)n}{1+n}\right]}\label{eq:Solve3}
\end{eqnarray}

\begin{eqnarray}
    \frac{dR}{dt} &=& \mu(N-R-S_0 e^{-\frac{\beta}{\mu}R\left[\frac{1+(1+\alpha k)n}{1+n}\right]})\label{eq:Solve4}
\end{eqnarray}

Solving the Eq. \ref{eq:Solve4}, we get
\begin{eqnarray}
    t &=& \frac{1}{\mu} \int_{0}^{R} \frac{dR}{N-R-S_0 e^{-\frac{\beta}{\mu}R\left[\frac{1+(1+\alpha k)n}{1+n}\right]}}\label{eq:Solve5}
\end{eqnarray}

As pandemic arrives at steady state when $t\longrightarrow \infty$ hence $\frac{dR}{dt}$ = 0 and $R_\infty$ = $constant$

\begin{eqnarray}
N-R_\infty = S_0 e^{-\frac{\beta}{\mu}R_\infty\left[\frac{1+(1+\alpha k)n}{1+n}\right]}\label{eq:Solve6}
\end{eqnarray}

Let initial conditions are $R(0) = 0$, $I(0)$ = $I$ and $S(0) = N - I \approx N$. Therefore, Eq. \ref{eq:Solve6} can be written as

\begin{eqnarray}
R_\infty &=& N - N e^{-\frac{\beta}{\mu}R_\infty\left[\frac{1+(1+\alpha k)n}{1+n}\right]}\label{eq:Solve18}
\end{eqnarray}

Normalizing the Eq. \ref{eq:Solve18}
\begin{eqnarray}
r_\infty &=& 1 - 1 e^{-R_0 r_\infty}\label{eq:Solve8}
\end{eqnarray}

Therefore, the reproduction number $R_0$ is
\begin{eqnarray}
R_0 &=& \frac{\beta}{\mu}\left[\frac{1+(1+\alpha k)n}{1+n}\right]\label{eq:Solve8}
\end{eqnarray}

In case there is no social connectivity to other locations ($\alpha = 0$ or $k=0$ or $n=0$) 
then the mobility 
SIR model will become the standard SIR model and the reproduction number is $R_0 = \frac{\beta}{\mu}$. Therefore, the reproduction number is directly proportional to social connectivity parameter $\alpha$, number of connected locations $k$ and depends upon individuals' mobility during a pandemic.

\begin{figure*}[ht!]
  \centering
  \subfloat[Susceptible Compartment]{\label{figur:11}\includegraphics[width=0.66\columnwidth]{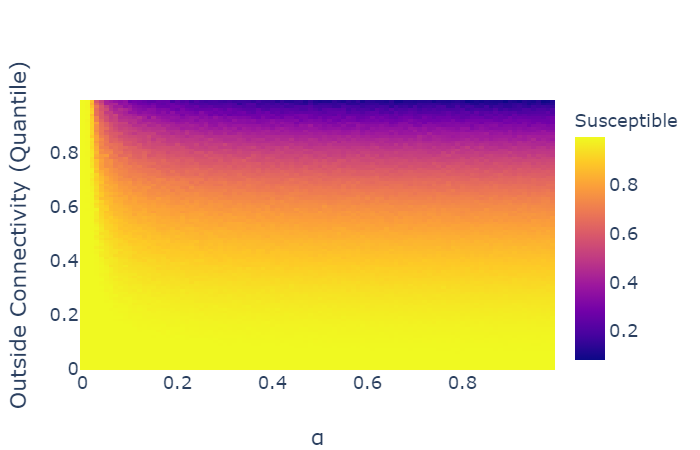}}
  \subfloat[Infected Compartment]{\label{figur:12}\includegraphics[width=0.66\columnwidth]{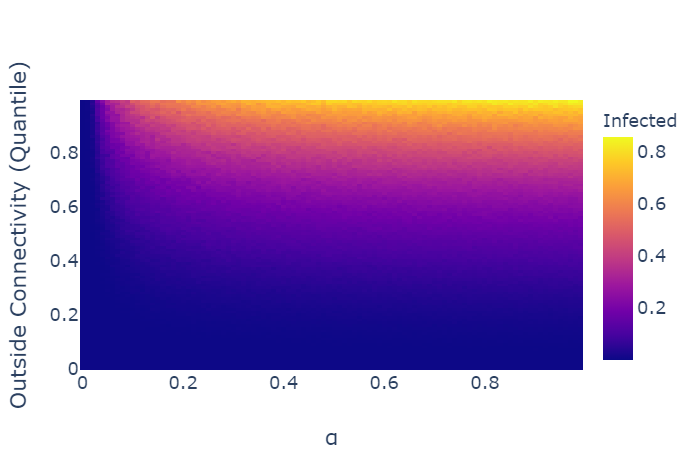}}
  \subfloat[Recovered Compartment]{\label{figur:13}\includegraphics[width=0.66\columnwidth]{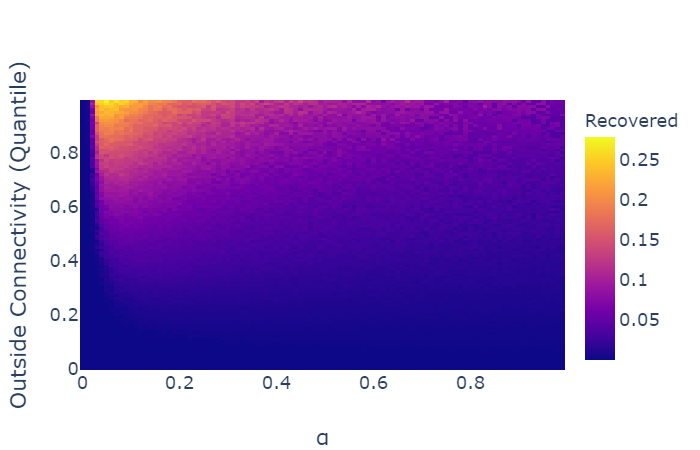}}
\caption{Pandemic Origin From Random Location: Numerical simulation of relationship between $\alpha$ and \textit{quarantine}}
\label{Fig:RelationRandom}
\end{figure*}

\begin{figure}[ht!]
    \centering
    \includegraphics[width=0.95\columnwidth]{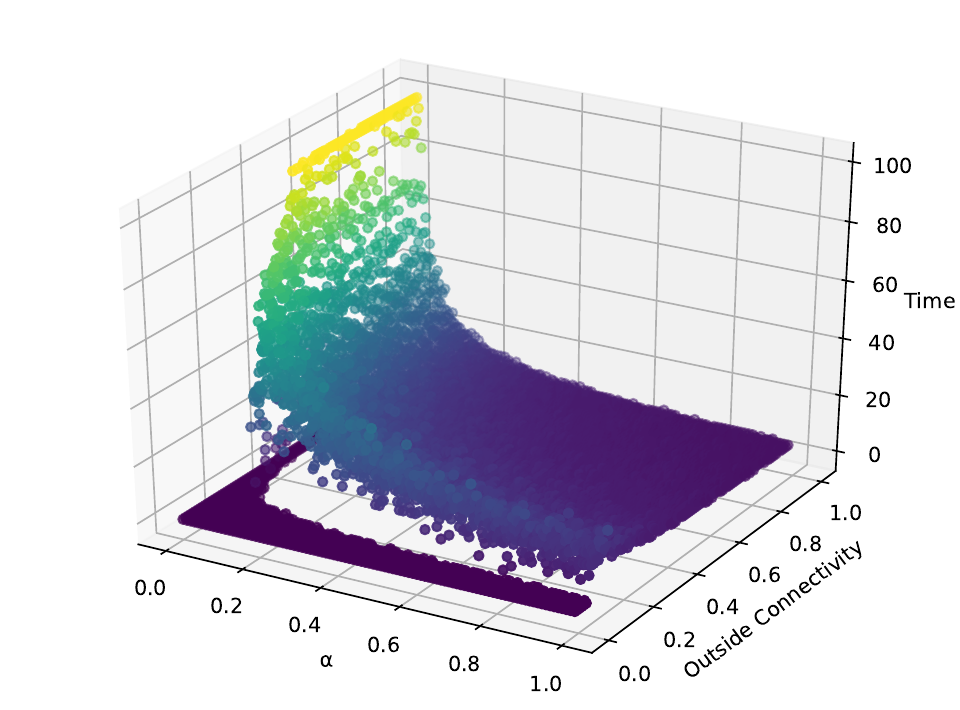}
    \caption{For different combinations of $\alpha$ and $quarantine$ percentile, number of days required to reach peak of infected compartment.}
    \label{fig:DaysCompare}
\end{figure}

\section{Evaluation} \label{sec:evaluation}
In this section, we first explain our experimental setup and next, we discuss the results of our simulation conducted using the proposed model 
on synthetic networks. In addition, we also applied our model for predicting the real-time Estonian COVID-19 cases.

\subsection{Experimental Setup}\label{subsec:ExpSetup}
For the analysis, we created an 
aggregated flow matrix of individuals per day from \textit{Origin to Destination (OD)}, which follows random distribution. Furthermore, three different techniques are 
considered for selecting the seed infection location:
\begin{enumerate}
    \item \textit{Pandemics origin from 
    a random location:} In this, a random location is 
    selected as seed infection location and a small fraction of individuals were infected at that location.
    \item \textit{Pandemics origin from a weakly connected location:} Here, seed location is selected strategically, which is weakly connected 
    to other locations. That implies least mobility of individuals from this location to other locations.
    \item \textit{Pandemics origin from a strongly connected location:} In this also, seed location is selected strategically, which is strongly connected 
    to other locations. This signifies that, highest mobility of individuals from this location to other locations.
\end{enumerate}

Our simulation is oriented towards addressing the following questions:
\begin{itemize}
    \item How \textit{social connectivity parameter} `$\alpha$' affects 
    the fraction of individuals in different compartments (\textit{susceptible, infected and recovered}) during a pandemic?
    \item What are the outcomes of 
    restricting the mobility (for top-X 
    percentile) of strongly connected locations?
    \item What is the relationship between \textit{social connectivity parameter} `$\alpha$' and the 
    mobility restriction (top-X percentile 
    of strongly connected locations?
    \item How efficiently this model can perform in real scenarios? 
    We answer this question by projecting the expected COVID-19 cases in Estonia.
\end{itemize}


\subsection{Results}\label{subsec:results}
We perform various 
simulation experiments to explain the proposed model on \textit{OD} network by using 
previously discussed techniques 
for selecting the seed infection location. 
It is to be noted that, if $\alpha =0$, then the model will behave as a standard SIR model. Also, if the mobility is reduced to 100 percentile (that is no mobility allowed) from strongly connected locations, then also model will act as a standard SIR model. 

\subsubsection{Pandemics Origin From a Random Location} 
Fig. \ref{Fig:AlphaRandom} displays the influence of the \textit{social connectivity parameter} 
'$\alpha$' while retaining the other parameters constant. Fig. \ref{figur:1} to \ref{figur:6} shows the pandemic dynamics with different values of $\alpha$ starting with $\alpha = 1$ to $\alpha = 0.1$. We observe that the peak of the infected compartment decreases significantly, as the $\alpha$ decreases, and it also takes longer to reach its peak. 
This indicates that there is a positive impact of lock-down in controlling a pandemic. 

The effect of 
restricting the mobility from the top-X 
percentile of highly connected locations with other locations is shown in Fig. \ref{Fig:QuantineRandom}. 
Fig. 
\ref{figur:7} to \ref{figur:10} displays the pandemic dynamics with different percentile of 
mobility restrictions of highly connected locations starting with 0\% to 30\% (keeping $\alpha = 0.5$). We observe that in case of pandemic, 
restricting the mobility from the top-10 percentile of highly connected locations 
can reduce the number of individuals 
who can get infected to 27\%. Therefore, quarantine 
plays a vital role during pandemics.

In order to understand the relationship between $\alpha$ and 
\textit{mobility restriction} from strongly connected locations, we performed the numerical simulation of the proposed mean-field equations (see Figure \ref{Fig:RelationRandom}). We can infer that \textit{social connectivity parameter} `$\alpha$' 
and 
\textit{mobility} both plays an important role during pandemics. Therefore, it is advisable to follow a dual strategy approach during a pandemic outbreak as 
\textit{controlling mobility} reduces the fraction of 
infected individuals and $\alpha$ delays the peak. 
Furthermore, we analysed the number of days required to reach 
the point where highest fraction of individuals get infected 
(see Figure \ref{fig:DaysCompare}). This indicates that 
mobility restrictions and minimal social contact will postpone the pandemic's peak and 
will give sufficient time for the preparations especially for the health sector. 

\subsubsection{Pandemics Origin From a Weakly and Strongly Connected Locations}
\begin{figure}[H]
  \centering
  \subfloat[Weakly connected ($\alpha$ = 1)]{\label{figur:14}\includegraphics[width=0.5\columnwidth]{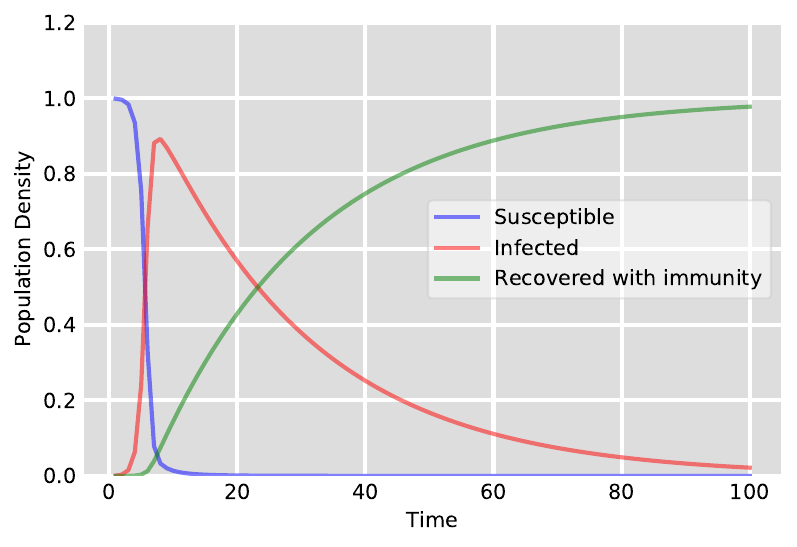}}
  \subfloat[Strongly connected ($\alpha$ = 1)]{\label{figur:15}\includegraphics[width=0.5\columnwidth]{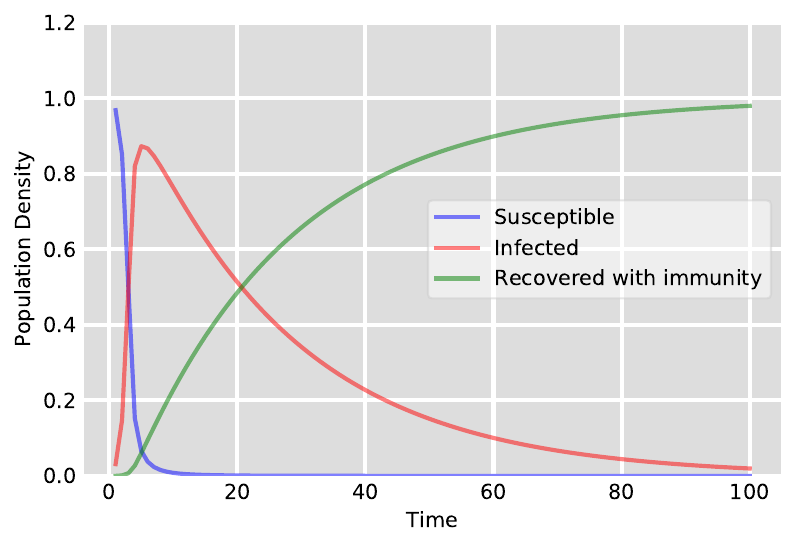}}
  \\
  \subfloat[Weakly connected ($\alpha$ =  0.8)]{\label{figur:16}\includegraphics[width=0.5\columnwidth]{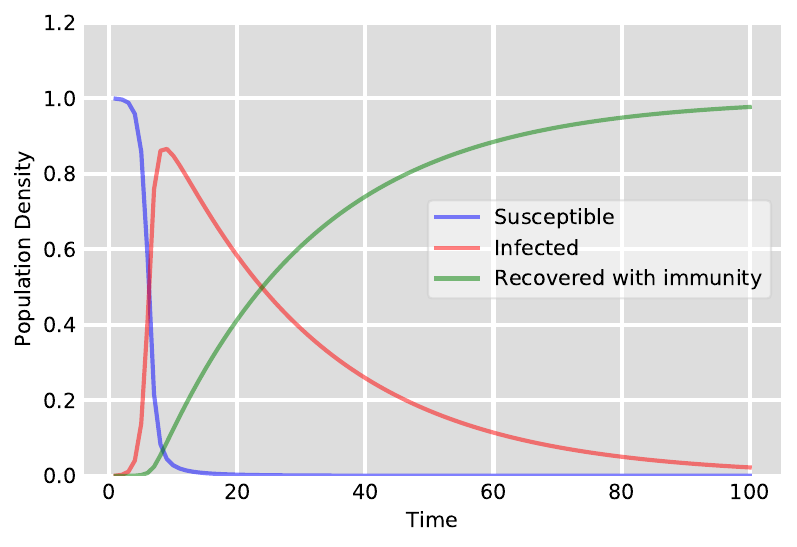}}
  \subfloat[Strongly connected ($\alpha$ =  0.8)]{\label{figur:17}\includegraphics[width=0.5\columnwidth]{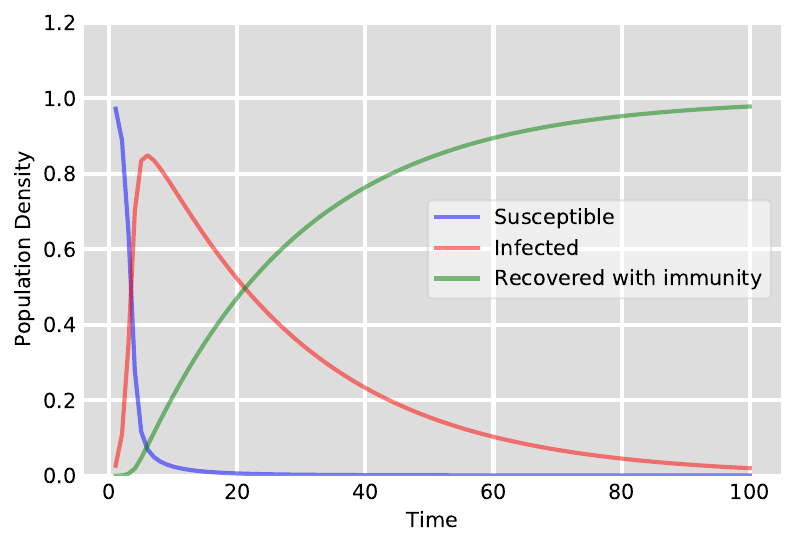}}
  \\
  \subfloat[Weakly connected ($\alpha$ =  0.5)]{\label{figur:18}\includegraphics[width=0.5\columnwidth]{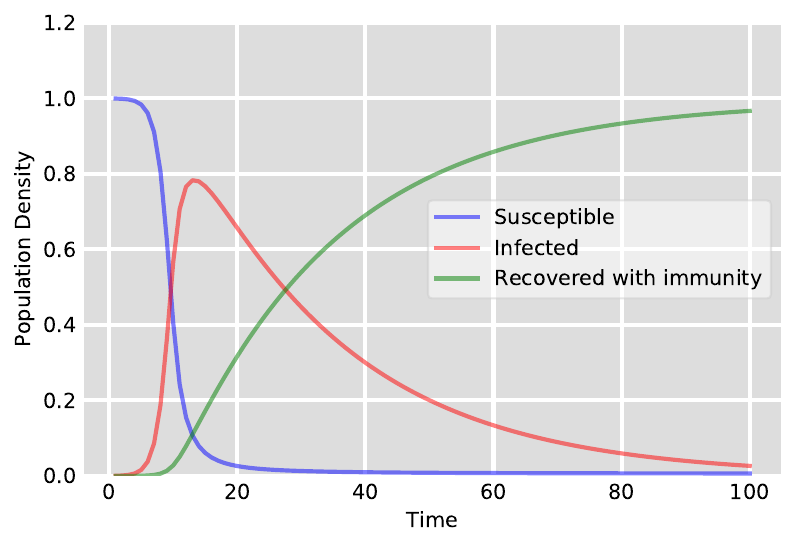}}
  \subfloat[Strongly connected ($\alpha$ =  0.5)]{\label{figur:19}\includegraphics[width=0.5\columnwidth]{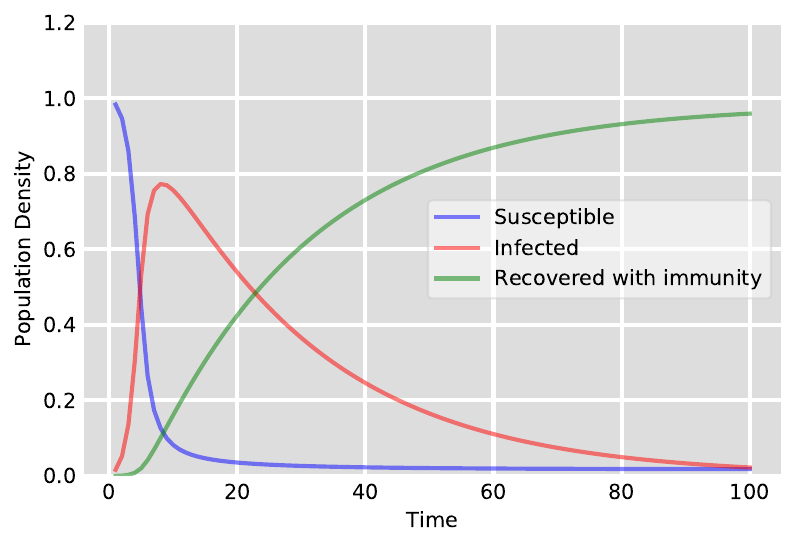}}
  \\
  \subfloat[Weakly connected ($\alpha$ = 0.4)]{\label{figur:20}\includegraphics[width=0.5\columnwidth]{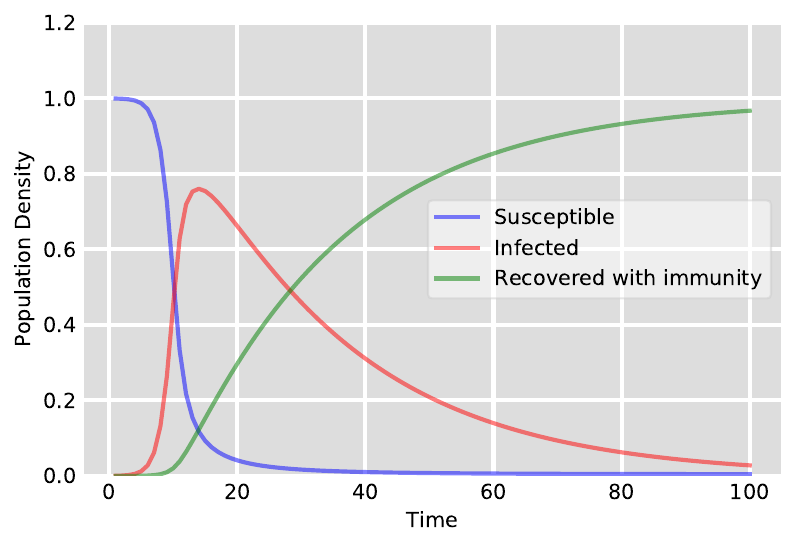}}
  \subfloat[Strongly connected ($\alpha$ = 0.4)]{\label{figur:21}\includegraphics[width=0.5\columnwidth]{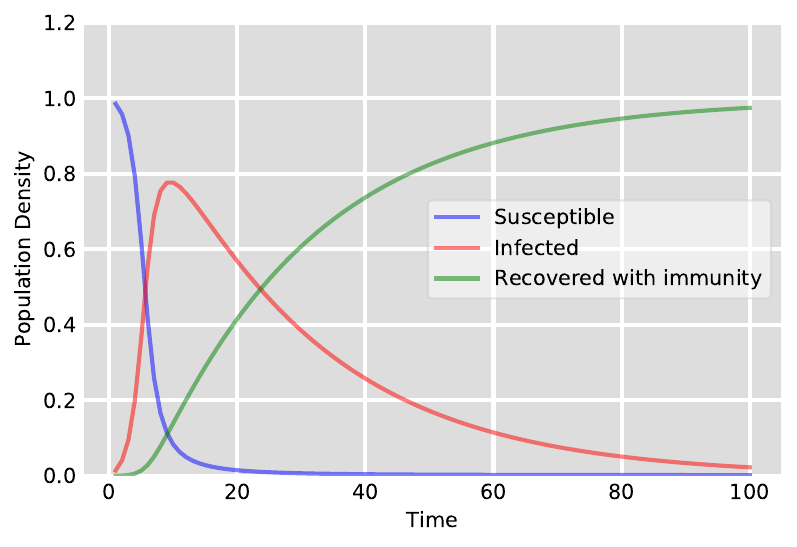}}
  \\
  \subfloat[Weakly connected ($\alpha$ =  0.2)]{\label{figur:22}\includegraphics[width=0.5\columnwidth]{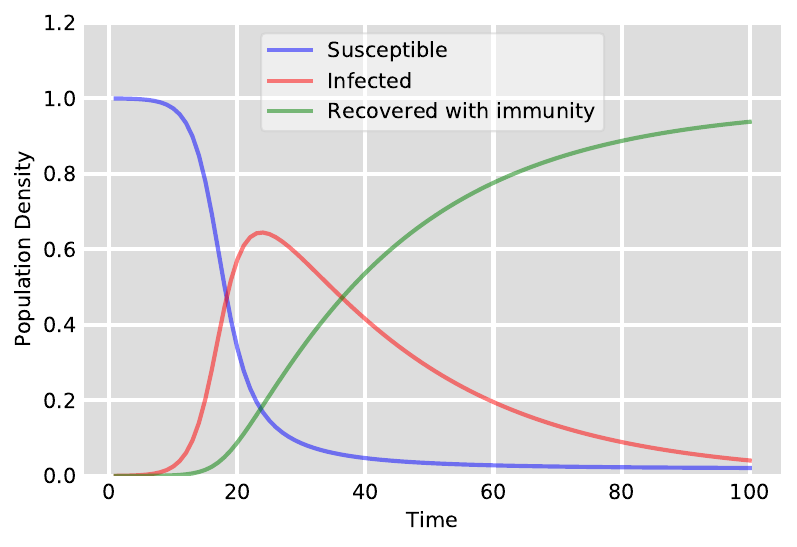}}
  \subfloat[Strongly connected ($\alpha$ =  0.2)]{\label{figur:23}\includegraphics[width=0.5\columnwidth]{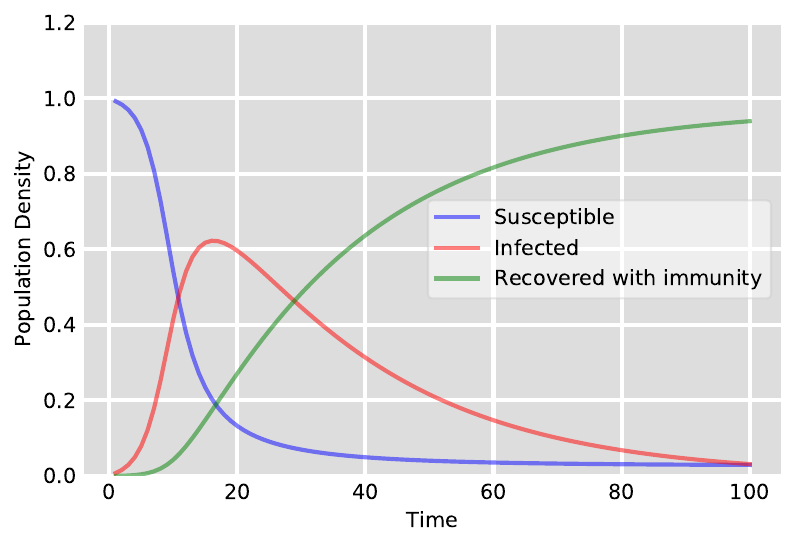}}
  \\
  \subfloat[Weakly connected ($\alpha$ =  0.1)]{\label{figur:25}\includegraphics[width=0.5\columnwidth]{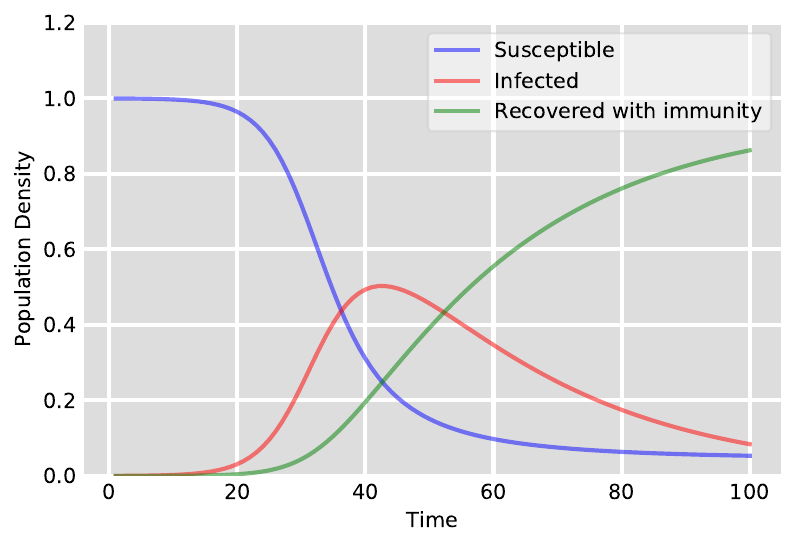}}
  \subfloat[Strongly connected ($\alpha$ =  0.1)]{\label{figur:26}\includegraphics[width=0.5\columnwidth]{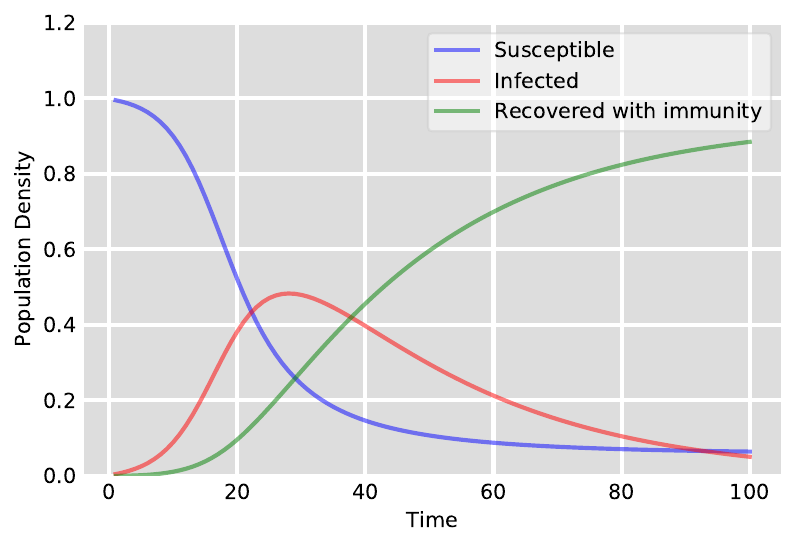}}
\caption{Pandemic Origin From Weak and Strongly Location: Effect of \textit{Social Connectivity Parameter} `$\alpha$'}
\label{Fig:AlphaWeakStrong}
\end{figure}

Fig. \ref{Fig:AlphaWeakStrong} displays the influence of the \textit{social communication parameter} '$\alpha$' while keeping 
the other parameters constant for both weakly and strongly connected locations. Fig. 
\ref{figur:14} to \ref{figur:26} shows the pandemic dynamics with different values of $\alpha$ starting with $\alpha = 1$ to $\alpha = 0.1$. 

It can be noted that when 
a pandemic originates from a weakly connected location, 
it takes longer to reach its peak compared to when 
it starts from a strongly connected location. This shows that location of origin also plays an important role during pandemic. Similar to random location, 
reducing mobility from the highly connected locations by 10 percentile can reduce the number of 
infected individuals 
between 18\% to 27\% for weakly and strongly connected locations. 

\subsubsection{Case Study Of Estonia} \label{subsubsec:Estonia}
\begin{figure}[ht!]
    \centering
    \includegraphics[width=\columnwidth]{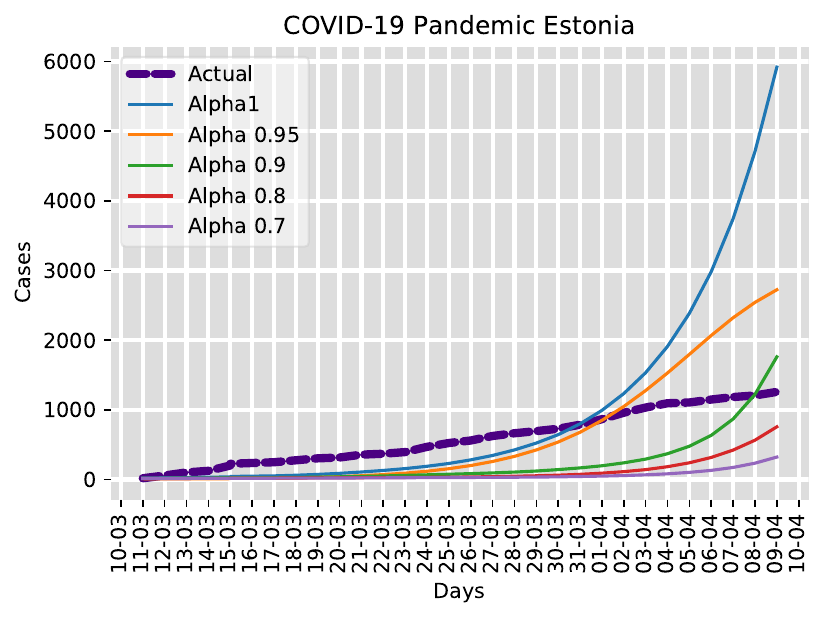}
    \caption{COVID-19 Cases In Estonia}
    \label{fig:Estonia}
\end{figure}

To 
demonstrate the usability of the model, we applied it on 
a real-time data of Estonia's {to fit} COVID-19 cases. Fig. \ref{fig:Estonia} shows the actual 
number of cases and 
the cases forecast by the model using different 
values for $\alpha$ 
and 
mobility percentile. For example, when $alpha = 0.95$, this indicates that social connectivity of individuals are reduced by 5\% and also top-5\% of strongly connected locations are 
restricted from mobility. Similarly, $\alpha = 0.7$, implies that social connectivity of individuals are reduced by 30\% and also the top-30\% of strongly connected locations have introduced restricted mobility.

For simulation, we created the $OD$ matrix between counties of Estonia using call data records \cite{hiir2019impact}. Furthermore, these call interactions are converted into population mobility between counties using 
Estonian population data \cite{estonia2018quarterly}. For the local transmission of the virus (within the county), we consider the reproduction number $R_0 = 2.5$ \cite{world2020coronavirus}.  

Cases reported until $11^{th}$ March, 2020  are considered as initial condition for the model. The reason behind selecting $11^{th}$ March, 2020 as initial condition is that, till this date no local transmission of the virus was reported\footnote{https://www.err.ee/1063204/terviseamet-eestis-on-kinnitatud-27-koroonajuhtu-ja-kohalik-levik}. Till the day of initial condition, the \textit{Estonian Health Board} confirmed 13 cases in \textit{Harju} and two cases in \textit{Tartumaa} and \textit{Saaremaa} 
each\footnote{https://www.terviseamet.ee/et/uuskoroonaviirus}. During the simulation, 
the number of cases in all other counties are initialized to 
zero. The infection rate $\beta$ and recovery rate $\mu$ are adjusted according to the value of $R_0$ for COVID-19. 
By $10^{th}$ April 2020, reported cases in Estonia and forecast cases 
using the model are 
shown in Fig. \ref{fig:Estonia}. It can be noticed that the model predicted much higher cases of COVID-19 if no restrictions are introduced ($\alpha$ = 1). However, as the restrictions were introduced by the Government\footnote{https://www.valitsus.ee/en/news} the number of cases got damped (Actual). Thus, the applicability of this model is to forecast a range of predicted number of cases which can help the governmental and health agencies to understand the impact and introduce proportional interventions to restrict the spread of the epidemic.
\section{Conclusion}\label{Sec:Conclusion}
Classical compartmental 
epidemic models are unable to describe the spreading pattern of pandemics such as COVID-19 
as they do not take into account the effect of \textit{social connectivity} and 
\textit{mobility} in spreading of 
the virus. 
Our proposed mobility based SIR 
model shows the 
significance of \textit{social connectivity} and 
\textit{mobility} during pandemics 
by 
taking into consideration the local and the global transmission rate of the 
infection. We have simulated the 
proposed model considering 
three different origins of the 
infection, 
namely random location, weakly connected location and strongly connected location. Our simulation shows that 
limiting the \textit{social connectivity} 
reduces 
and delays the peak of the infected compartment. 
Our analysis also shows that 
restricting the mobility from the top-10 percentile of connected locations can reduce the number of 
infected individuals 
between 18\% to 27\%. From the mathematical proof for our proposed model, we obtained that the reproduction number $R_0$ directly depends upon \textit{social connectivity of individuals}, \textit{number of connected locations} and \textit{individuals mobility between locations} which is in line with our simulations' results. 
This indicates that introducing \textit{isolation} and \textit{quarantine} is effective in fighting a pandemic crisis.  Using the 
proposed model, we also simulated the  real world scenario by considering the COVID-19 cases in Estonia. Simulation reveals that 
the mobility based SIR model 
can be helpful to forecast the expected number of cases after some proportion of \textit{isolation} and \textit{quarantine} is introduced in the society.

We plan to include various future directions for this work 
such as by simulating the model using additional dynamic networks.  
Another direction would 
be to use 
additional mobility data 
such as transportation network for better understanding the pandemic behavior. Importantly, we plan to introduce infection delay and recovery delay simultaneously in our future studies. 

\section*{Acknowledgment}
This research was funded by ERDF via the IT Academy Research Programme and SoBigData++.

\bibliographystyle{plain}
\bibliography{main}

\end{document}